# Silicon Mie Resonators for Highly Directional Light Emission from monolayer MoS$_2$

Ahmet Fatih Cihan[1], Alberto G. Curto[1,2], Søren Raza[1], Pieter G. Kik[1,3]

and Mark L. Brongersma[1]

1. Geballe Laboratory for Advanced Materials, Stanford University, Stanford, California 94305, United States
2. Dep. Applied Physics and Institute for Photonic Integration, Eindhoven University of Technology, 5600 MB Eindhoven, The Netherlands
3. CREOL, The College of Optics and Photonics, University of Central Florida, Florida 32816, United States

**Abstract:**

Controlling light emission from quantum emitters has important applications ranging from solid-state lighting and displays to nanoscale single-photon sources. Optical antennas have emerged as promising tools to achieve such control right at the location of the emitter, without the need for bulky, external optics. Semiconductor nanoantennas are particularly practical for this purpose because simple geometries, such as wires and spheres, support multiple, degenerate optical resonances. Here, we start by modifying Mie scattering theory developed for plane wave illumination to describe scattering of dipole emission. We then use this theory and experiments to demonstrate several pathways to achieve control over the directionality, polarization state, and spectral emission that rely on a coherent coupling of an emitting dipole to optical resonances of a Si nanowire. A forward-to-backward ratio of 20 was demonstrated for the electric dipole emission at 680 nm from a monolayer MoS$_2$ by optically coupling it to a Si nanowire.

**Main text:**

Achieving control over the radiation properties of quantum emitters is key to improving efficiency and realizing new functionality in optoelectronic systems. Bulky optical components have been developed for many years and are extremely effective in controlling the angular, polarization, and spectral properties of light emission. Recent advances in the fields of metallic and dielectric optical metamaterials and nanoantennas have now also enabled effective integration of solid-state emitters and control elements into inexpensive platforms.[1–3] Such structures can manipulate light emission in the near-field of an emitter and thus hold a real promise to achieve even greater control over the emission process. For example, we will show how the undesired losses due to radiation of quantum emitters into a high-index substrate can be reduced by redirecting the emission upward with an antenna.

Whereas structures based on noble metals are currently most advanced in manipulating light-matter interaction at the nanoscale, they typically are complex in shape, display undesired optical losses, and are not compatible with most semiconductor device processing technologies. High-index semiconductor antennas can circumvent these issues while providing complex electrical and optical functions.[2,4–14] Based on the mature fabrication infrastructure, silicon nanostructures appear particularly promising for optoelectronic applications.[4,9–12,15–17] Semiconductor nanoparticles of simple geometric shapes have displayed directional scattering of plane waves when the renowned Kerker conditions are satisfied.[12,16,18] When these conditions are met, directionality is naturally achieved through the coherent excitation of electric and magnetic dipole resonances in the particle and tuning the interference of the associated scattered fields.[12,19,20] Thanks to their high refractive indices, semiconductor nanoparticles can satisfy the Kerker conditions in the visible spectral range.[16,18,21] Given the ever-increasing importance of solid state light emitters and quantum nanophotonics, it is of great interest to explore whether analogous conditions can be identified that will facilitate directional emission from quantum emitters and we answer this is important question positively in this work. As such, it nicely complements other low-loss approaches involving advanced semiconductor photonic crystals and leaky wave antenna structures to control spectral and angular emission properties.[22–27]

Directional emission with the help of nanometallic antennas has been analyzed theoretically in great detail and was demonstrated experimentally at optical frequencies.[28–33] For semiconductor antennas, however, directional emission exploiting Mie resonances has been limited to theoretical proposals[16,34–46] or experiments in the microwave regime.[47] By modifying the conventional Mie theory to describe light scattering by a nanowire (NW) from a dipolar source as opposed to the standard plane-wave source, we reveal that directional emission with a silicon NW can be realized through a variety of mechanisms. Each of these directionality mechanisms involves optical interference effects that come about when the light emitted from a quantum emitter can follow different pathways to the far-field. For example, highly-directional emission can occur when a fraction of the light emitted by an electric dipole source is rescattered by exciting the dominant electric dipole resonance of a nearby NW. It can also follow from the coherent excitation of electric and magnetic dipoles in a NW, more akin to the original directional Kerker scattering.[38] Experimental evidence is provided for both of these directionality mechanisms by coupling the exciton emission from an atomically thin layer of $MoS_2$ [48–52] to two differently-sized Si NWs. We demonstrate a forward-to-backward emission ratio of 20 from $MoS_2$ emitters coupled to Si NWs at visible wavelengths. The use of a 2D semiconductor enables the realization of an accurate and repeatable separation between the emitter and NW, which is critical for obtaining reliable control over the direction of light emission.[37] Furthermore, 2D materials such as $MoS_2$ benefit from the simplicity and low-cost nature of available deposition/growth techniques making them particularly promising for future photonic device applications compared to conventional epitaxially-grown emitter materials.[53–56]

Conventional Mie theory allows us to calculate the excited resonances for an infinitely long cylinder under plane-wave illumination (See Fig. 1a for the excitation geometry) by expanding the incident and scattered fields in terms of cylindrical harmonics.[57] The expansion coefficients in the scattered field are referred to as Mie coefficients and they quantify the scattering contributions from different resonances, e.g. electric and magnetic dipoles (see Figure 1b for a multipolar decomposition of the scattering efficiency $Q_{\text{scat}}$ for an infinitely long Si NW). When Mie coefficients corresponding to the electric and magnetic dipoles excited by a plane wave have equal amplitude and phase, the scattered fields originating from these resonances destructively interfere in the backward direction, satisfying the so-called 1$^{\text{st}}$ Kerker condition.[19] This condition has played a central role in the research on light-matter interactions with high-index nanostructures.[3] In this work we will go beyond plane wave excitation and explore the impact of using the electric dipole emission from the MoS$_2$ monolayer as our source. We show that the choice of a localized source can profoundly influence the weights of the excited NW resonances compared to the plane-wave case. This naturally results in different conditions for emission directionality. To gain an intuitive understanding of the origin of these modifications, we derive an analytic expression for the scattering efficiency of a cylindrical NW for electric dipole radiation in two dimensions (2D):

$$Q_{\text{scat}}^{\text{dipole}} = \frac{P_{\text{scat}}}{P_{\text{inc}}} = \sum_{n=-\infty}^{\infty} |b_n|^2 \left| H_n^{(1)}[k_0(R+d)] \right|^2. \quad (1)$$

Here, $P_{\text{scat}}$ and $P_{\text{inc}}$ are the scattered and incident source-dipole radiation powers respectively, $b_n$ are the conventional Mie coefficients found from the plane-wave excitation problem,[57] $H_n^{(1)}$ are Hankel functions of the first kind, $k_0 = 2\pi/\lambda$ is the free-space wave vector, $R$ is the NW radius and $d$ is the dipole-NW distance (see Supplementary Information for the derivation of Eq. 1). Note that in this 2D analysis, the excitation source is a line dipole and the NW is infinitely long.[57,58] Whereas a 3D theory would be needed to quantitatively predict the correct magnitude for the scattering efficiency for a point dipole source, we show that this simple 2D description can explain the role of the different multipolar excitations in achieving directional emission and the importance of controlling the NW-emitter distance.[37] This is particularly true as, for symmetry reasons, the far-field radiation patterns seen for an infinitesimal electric point-dipole (3D) and an electric line-source (2D) are identical in the plane orthogonal to the NW-axis and intersecting the emitter. (See Supplementary Fig. S2). Hence, the scattered far-field radiation pattern in this plane will also be identical in both cases.[45]

As can be seen in Eq. 1, each term in the expression for the scattering efficiency can be associated with a specific resonance. Each of the resonant contributions has a distance dependent weighting factor that is given by the Hankel function relevant to that resonance. The origin of this Hankel function dependence arises as the scattered power related to the interaction of the dipole source and NW scattered fields, which are both described by Hankel functions. Because of these weighting factors, the excitation efficiencies for different multipoles display different dependences on the dipole-NW distance, which makes this distance an important parameter determining the directionality of the overall emission. The scattering behavior of the NW under dipole illumination approaches that seen for plane wave illumination when the dipole source is far away ($d \gg \lambda$) from the NW (Supplementary Figure S1). However, when the dipole is brought closer to the NW as in Fig. 1c, the excitation of different resonances takes place with different strengths resulting in a modified scattering spectrum (Fig. 1d).

The dipole emission directionality (Fig. 1c) is different from the plane wave scattering directionality not only because of the modified resonance contributions but also because of the non-directional nature of the source.[38] In other words, to achieve the desired emission directionality, it is important to consider the interference of the incident fields from the source and scattered fields from the NW as opposed to the plane wave case where typically only the scattered fields are considered.[16,18,21]

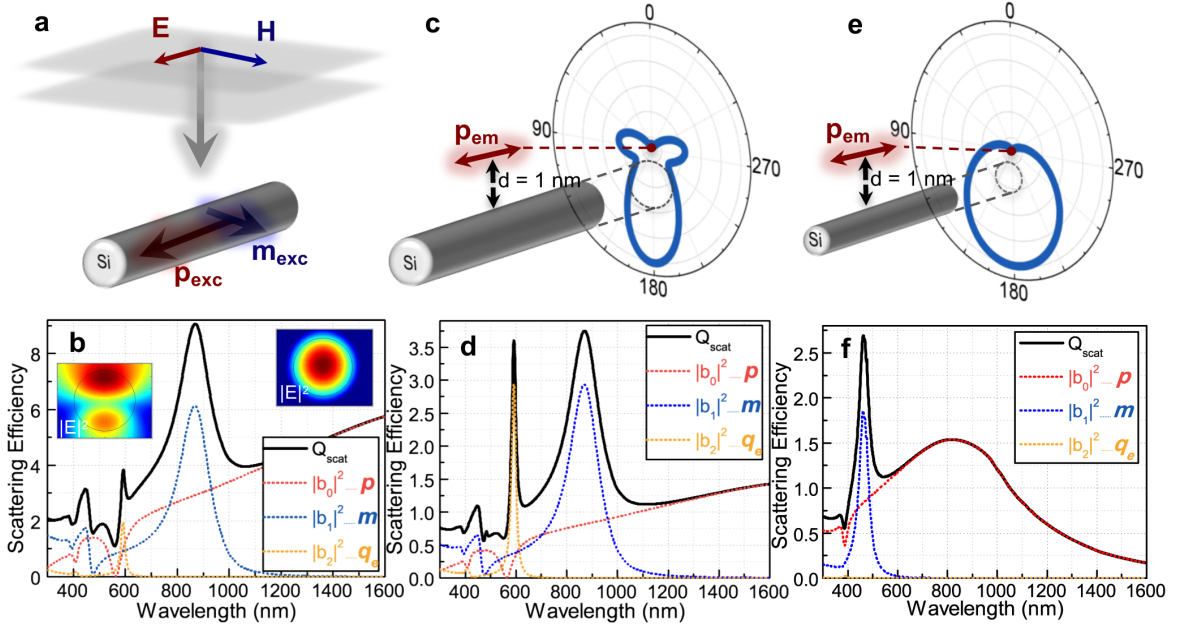

*Figure 1. Scattering behavior of a silicon nanowire under plane wave and dipole excitations. a, Schematic of the NW under plane wave excitation showing the possible excitation of electric and magnetic dipole resonances. b, Scattering efficiency and the multipolar contributions from the electric dipole p, magnetic dipole m and electric quadrupole $q_e$ resonances under plane wave excitation for a NW with an 88 nm radius. Insets are the electric field intensity profiles in the NW at the electric (right) and magnetic (left) dipole peak wavelengths. Schematic and the radiation pattern for dipole excitation of a NW-radius of c, 88 nm or e, 38 nm and a dipole-NW distance of 1 nm at a wavelength of 675 nm. d and f, Scattering efficiency and its multipolar contributions for cases corresponding to c and e, respectively.*

For a given emitter/emission wavelength, the NW radius can be optimized in a thoughtful manner to achieve directional emission; as it controls the nature of the excited resonances, it effectively controls the scattering behavior. By using NWs with different radii, one can obtain directionalities with distinct origins even at the same emission wavelength, as seen in Figs. 1c and 1e. With the larger radius NW shown in Fig. 1c, higher-order multipoles can be excited in the emission band of MoS$_2$ in the range from 650-720 nm, resulting in directionality due to the interference of the scattered fields originating from these multipoles and the emitter dipole. On the other hand, with the smaller NW of Fig. 1e, we can observe directional emission due to the interference of the scattered fields from the excited electric dipole NW resonance with those of the emitter dipole.

To experimentally investigate the emission directionality, we use a confocal optical microscope with a laser excitation wavelength of 633 nm that affords both top and bottom detection of photoluminescence (PL) signals (see Methods). As the emitter, we use monolayer MoS$_2$ grown on a transparent sapphire substrate. Si NWs are drop-cast on top, as shown in Fig. 2a. In one experiment, we use a tapered Si NW whose radius gradually varies from around 20 nm to 40 nm over a length of 40 μm. The gently tapered shape of the NW makes it convenient to match the spectrum of the emitter to the different diameter-dependent NW resonances that give rise to directionality.

Images of the transverse magnetic (TM) polarized emission are taken from the top and bottom of the sample (Fig. 2b,c). They show that the NW enhances the emission of the MoS$_2$ layer to the top (i.e forward-direction) while suppressing the emission to the bottom (i.e. backward direction). The top-to-bottom (T/B) ratio for the emission, provided in Fig. 2d, reaches values up to 2.5 on the NW while that of bare MoS$_2$ areas is about 0.8 (see Methods for the procedure to correct for the contribution of the detection system to the measured asymmetry). An important caveat of the measurement in Fig. 2d is that the detection system is diffraction-limited, and consequently the region from which photoluminescence is collected is significantly larger than the NW diameter. This results in a reduced T/B ratio compared to our simulations due to the contributions from the bare MoS$_2$ regions adjacent to the NW which produce non-directional emission. In order to eliminate this unwanted effect and measure the ultimate directionality observable for emitters under the antenna, the sample is etched using a low-energy Ar plasma where the NW itself acts as an etching mask (see Fig. 2e).

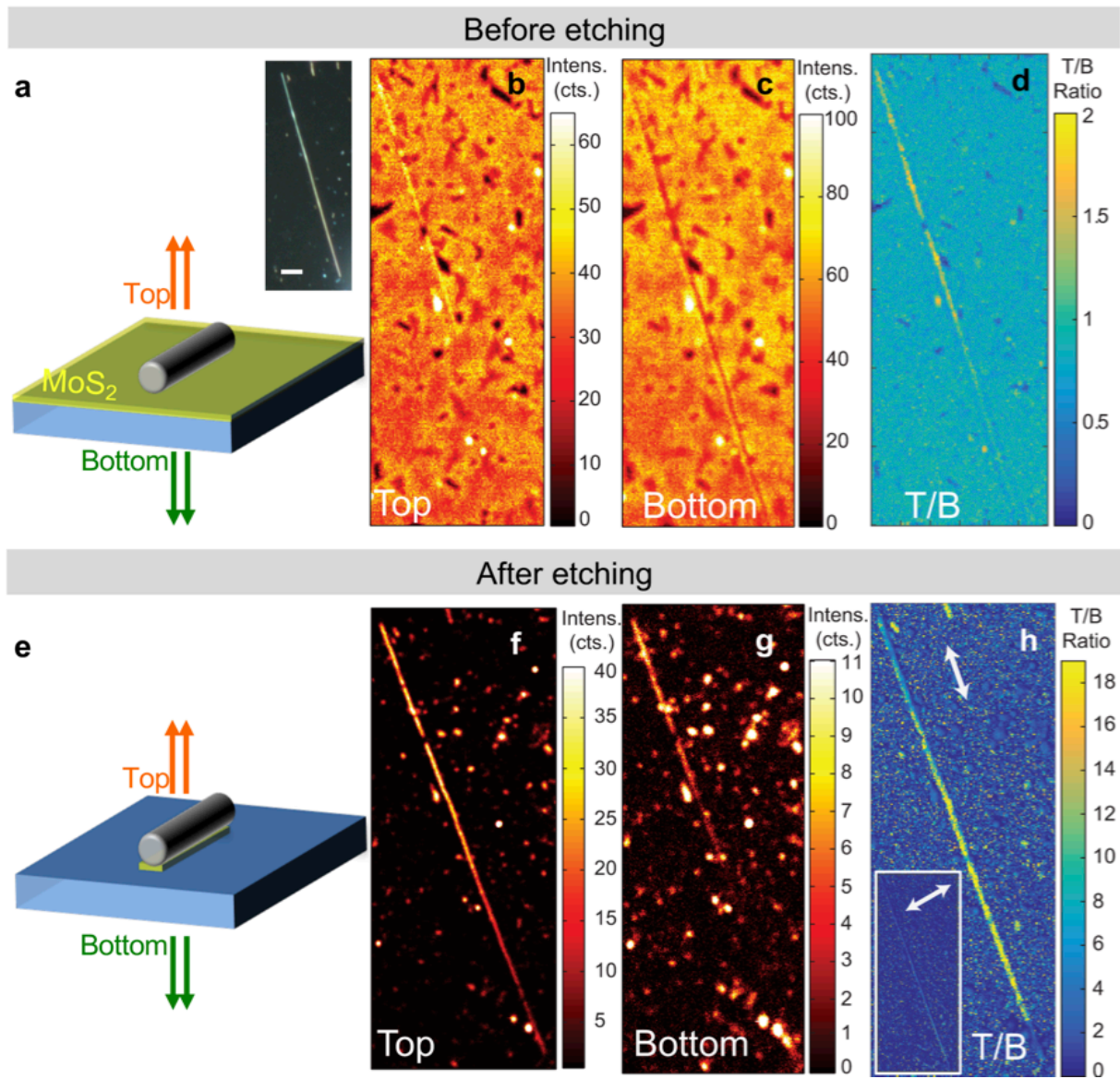

*Figure 2. Experimental demonstration of highly directional emission in fluorescence images. **a**, Schematic and dark-field scattering image of an un-etched MoS$_2$ sample. The scale bar corresponds to 5 μm. **b, c,** Top (T) and bottom (B) fluorescence images of the sample for TM polarization. **d,** T/B ratio normalized such that the bare MoS$_2$ T/B ratio is ca. 0.8, as predicted by optical simulations for this elementary system (see Methods). **e-h,** Corresponding images for the sample after etching the MoS$_2$ around the NW. Inset to panel **h** shows that the T/B ratio for the TE polarization displays no noticeable enhancement. The arrows indicate the electric field polarization direction for the collected light.*

The top and bottom fluorescence images after the etching process show that the emission is almost entirely eliminated from the bare $MoS_2$ regions, except for isolated spots corresponding to areas of few-layer $MoS_2$, while the emission from the emitters under the NW is maintained (Fig. 2f, g). Note that the bottom emission around the lower half of the NW in Figure 2g is almost completely suppressed, suggestive of a very high directionality (see Supplementary Information for top and bottom line scans of the emission across the NW). As can be seen in Figure 2h, the T/B ratio is more than 20 on the part of the tapered NW whereas that of the reference bare $MoS_2$ is around 0.8. Therefore, it can be concluded that the NW directionality enhancement over bare $MoS_2$ is more than 25-fold.

As a control experiment, we provide the T/B ratio of the same NW for TE polarized collection in the inset of Figure 2h. Here, the NW is barely discernable in the T/B ratio image implying that the enhanced directionality seen for the TM case is linked to multipolar resonances capable of redirecting the emission with this polarization state.

In order to explore the nature of the observed directional emission, we use our analytical Mie theory to calculate the T/B emission ratio. Within this approach, the far-field emission intensity is given as:

$$I_{ff}(\theta) = \frac{\omega^2 \mu_0^2 I^2}{8\pi} \left| \sum_{n=-\infty}^{\infty} (-i)^n \left\{ J_n[k_0(R+d)] - b_n H_n^{(1)}[k_0(R+d)] \right\} e^{-in\theta} \right|^2 \quad (2)$$

where $J_n$ is the Bessel function of the first kind, $I$ is the current of the line source, and $\theta$ is the emission angle (see Supplementary Information for the derivation). By integrating the far-field intensity values in Eq. (2) around the forward ($\theta = \pi$) and backward ($\theta = 0$) directions, we obtain the spectral and NW size dependence of the T/B emission ratio (see Methods). Figure 3a shows a map of these calculations, revealing multiple clear bands with high T/B ratios. Each band can be identified with a unique directionality mechanism involving different multipolar resonances. Very similar results are obtained with FDTD simulations that include the presence of the substrate, as shown in supplementary Fig. S4.

For the emission wavelength range of MoS$_2$ (the vertical shaded area in Fig. 3a), it is possible to capture two directionality mechanisms with NWs that feature radii below 100 nm.

In order to identify how the different multipoles contribute to each mechanism, we determine the far-field complex electric field contributions in the backward direction from several multipolar resonances. We first analyze the smaller, tapered NW for which the high directionality was verified in Fig. 2. Its radius range from 20 nm – 40 nm is indicated by the lower grey-shaded band. For a selected 38 nm radius NW, these relevant field contributions are shown in the left panel of Fig. 3b in a phasor representation to clearly show their relative magnitudes and phases. It is clear that for this wire the excitation of an electric dipole ($p_{exc}$) resonance alone suffices to almost fully cancel out the incident field originating from the source dipole in the backward direction. There is only a very small contribution related to the magnetic dipole $m_{exc}$. However, for a larger 88 nm-radius NW (right panel of Fig. 3b), fields from the electric dipole, magnetic dipole and electric quadrupole resonances ($p_{exc}$, $m_{exc}$ and $q_{exc}$, respectively), are more-or-less equal in magnitude. This corresponds results in very little backward emission and in this case we find that the directionality can primarily be attributed to the coherent excitation of electric and magnetic dipole resonances. The excitation of these resonances result in scattered fields that cancel the field originating from source dipole ($p_{inc}$) in the backward direction. Figure 3c shows experimental evidence for this directionality mechanism. For this 88-nm-radius wire , a T/B ratio in excess of 5 is reached. The lower T/B ratio observed for this larger wire is consistent with the theoretical analysis shown Fig. 3a.

As with any antenna the scattered field can produce a back action on the dipole source and an associated Purcell enhancement.[59] As the relative strengths of the source and scattered fields control the directionality, it is important to be aware of such effects. As our model is an exact solution to Maxwell's equations, the impact of possible Purcell enhancements due to the presence of the NW are naturally accounted for.

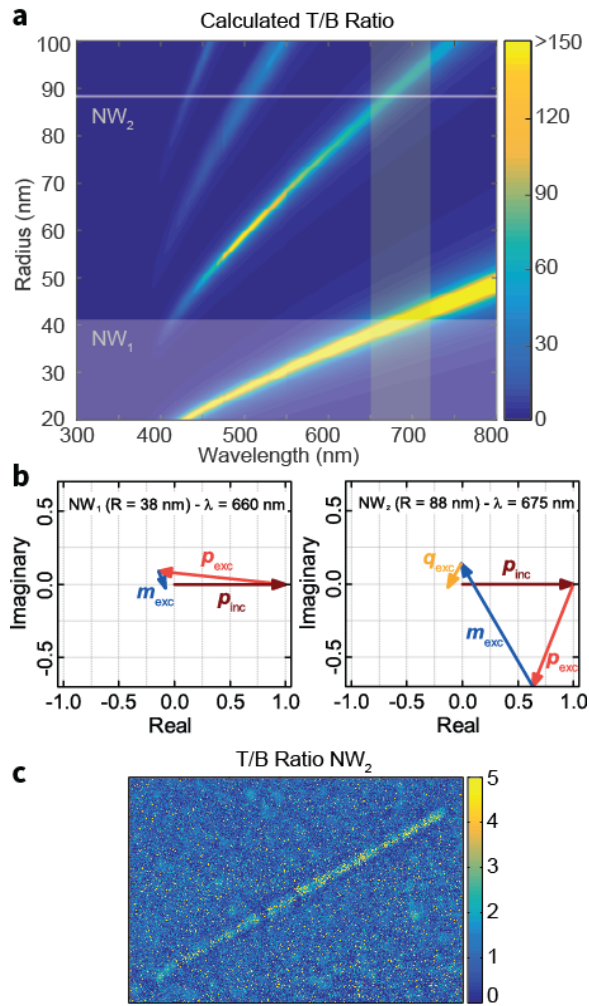

*Figure 3. Analysis of directionality mechanisms and experimental demonstration of the second mechanism. a,* T/B ratio of emission of a dipole-NW pair obtained with the modified Mie theory (see Methods). The vertical shaded area corresponds to the emission band of $MoS_2$ and the horizontal line/shaded band indicate the sized of the two NWs used in this study. *b,* Phasor representations of the electric fields at a position in the backward emission direction and broken down into different multipolar contributions for NWs with radii of 38 nm ($NW_1$, left panel) and 88 nm ($NW_2$, right panel). Labels indicate the sources of each contribution. The electric field due to the source dipole is taken to be 1 and the other components are normalized accordingly. *c,* Map of the experimental T/B ratio for the 88 nm radius NW, obtained in the same way as in Fig. 2.

Besides the demonstrated angular and polarization control, it is also highly desirable to achieve spectral control over the emission right at the source. This can be attained through a simple variation in the NW size.[11,60] For example, Fig. 3a suggests that for each directionality mechanism, the maximum in the T/B ratio will redshift with increasing NW size. This is understandable as all of the NW resonances that may be involved in achieving directionality redshift with increasing size. At wavelengths where the T/B ratio is large, it is expected that a relatively large fraction of the $MoS_2$ PL can be collected in the top direction. It is thus also expected that the PL spectra collected from the top will redshift as the NW size increases. To investigate this point, we analyze a zoomed-in top PL emission map (see Fig.4a) from the tapered NW that was discussed in Fig. 2. Reflection mode dark-field white light scattering spectrums taken from different locations along the length of the NW show a clear redshift over about 60 nm as the radius increases from 26 nm to 38 nm (see Fig. 4b inset). This comes with a concomitant redshift in the PL spectrum, as shown in Fig. 4c. As the detection spot is moved along the NW length, the emission spectrum is first higher on the blue-side of the reference bare $MoS_2$ emission peak and ultimately the peak is on the red-side of this spectrum (Fig. 4c).

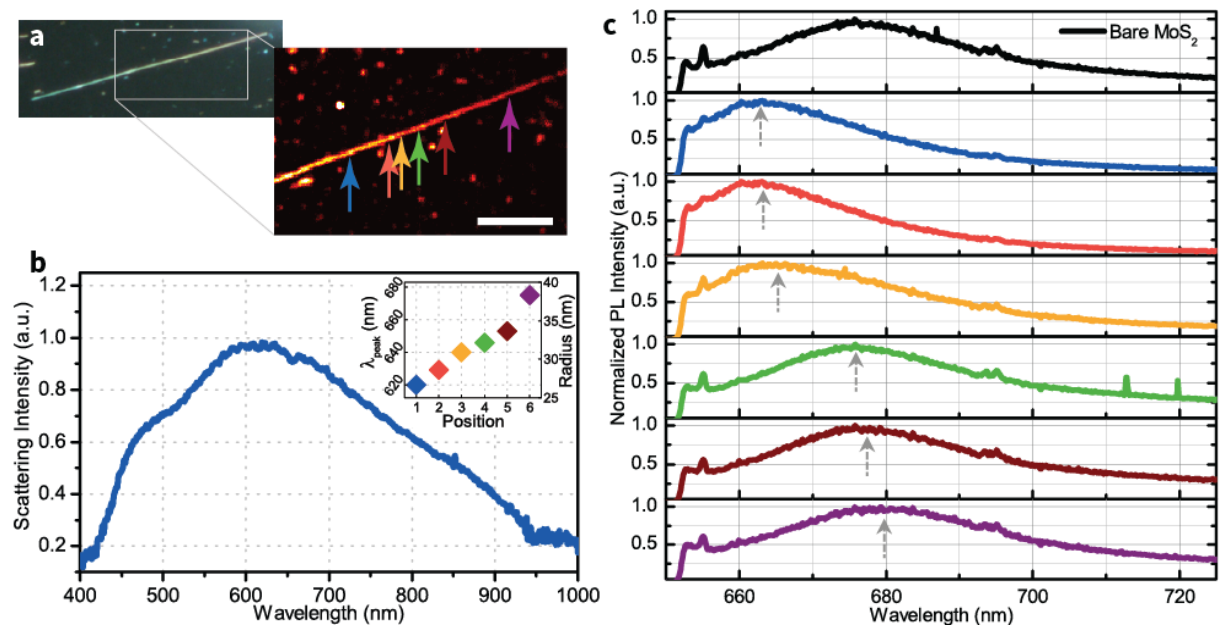

*Figure 4. Spectral control over the emission wavelength with NW size. a,* Zoom-in of the top fluorescence image of the Si NW that is tapered from 20 nm (left side) to 40 nm (right side). The colored arrows indicate the spatial locations along the NW where the dark-field white light scattering and PL emission spectrums are taken. The scale bar corresponds to 4 μm. *b,* Dark-field scattering spectrum of the NW for TM mode collected confocally from the spot on the NW corresponding to the blue arrow in part a. Inset shows the evolution of the scattering peak and the radius obtained by atomic force microscopy as the collection point is moved from left to right on the NW. The colors of the data points match the colors of the lines in part a and c. *c,* Emission spectra of the system across the NW. Each colored curve corresponds to a position indicated with the same-color line in part a. Top black curve is the reference $MoS_2$ emission spectrum taken from a bare $MoS_2$ region on the sample. Grey arrows are guides to the eye to visualize the shift in the PL peak emission wavelength.

In conclusion, we have experimentally demonstrated directional and spectral control over the visible emission from monolayer $MoS_2$ emitters with Si nanowires. A more than 25-fold top-to-bottom emission ratio enhancement was observed indicating that dielectric antennas can be very effective in reducing the undesired emission from quantum emitters into high-index substrates. Conventional Mie theory was extended to the use of an electric dipole source to explain the observed directionality in terms of multipolar contributions with an intuitive analytical model. The insights from this work may find use in realizing high-performance single-photon sources or metasurfaces composed of dense arrays of high-index semiconductor nanostructures to enhance and control light extraction from solid-state light emitters.

## Methods

**Sample Preparation and NW properties.** The tapered NW discussed in Figs. 2-4 with the radius in the range of 20-40 nm, was grown in-house by a gold-catalyzed chemical vapor deposition technique.[6,61] The second NW of this study, $NW_2$ with a radius of 88 nm was purchased from Sigma-Aldrich Co. $MoS_2$ monolayer samples with semi-continuous areas, grown by chemical vapor deposition, were purchased from 2D Semiconductors Inc. The granularity of PL maps in the paper is due to this imperfect surface coverage of $MoS_2$. The NWs were drop-casted on these $MoS_2$ samples from a colloidal suspension.

**$MoS_2$ layer etching process.** In order to eliminate the collection of emission from the $MoS_2$ areas that are not optically coupled the NWs, we expose the sample to a mild Ar-plasma etching process for 10 seconds in a sputtering system (AJA International Inc.). The NWs themselves serve as a shadow mask protecting the emitter layer underneath while the rest of the $MoS_2$ monolayer is removed.

**Photoluminescence imaging and spectroscopy.** Experiments were conducted on a modified confocal microscope with top and bottom detection capabilities (Witec Alpha 300). The sample was excited with 633 nm continuous-wave laser through the top objective. The top detection path consists of an objective (Zeiss 100X, NA: 0.95), a linear polarizer, a laser filter (long-pass with 650 nm cutoff), a fiber with a 20 μm core diameter serving as the confocal pinhole and a fiber-coupled avalanche photo-diode (Micro Photon Devices (MPD)). The bottom detection path consists of an objective (Zeiss 50X, NA: 0.55), a linear polarizer, a laser filter (long-pass with 650 nm cutoff), and a free-space coupled avalanche photo-diode (MPD). The top objective was chosen to have a high numerical aperture (NA) to increase the excitation resolution of the system. The bottom objective, on the other hand, was a long working distance objective to allow imaging through the transparent substrate and hence had a lower numerical aperture. T/B ratio is calculated after a background subtraction from the top and bottom images.

**FDTD simulations for reference and normalization procedure for Top/Bottom ratio analysis.** Because of the inherent asymmetry of the top-bottom detection system (different objectives, free-space vs fiber-coupled detection, beam-splitter in the top direction for laser coupling, and the sample asymmetry due to substrate), a reference sample is needed to obtain a meaningful value for the T/B ratio. We used the T/B ratio of bare $MoS_2$ emission on sapphire substrate as a relevant reference. For this simple geometry the T/B ratio is easily determined by 3D FDTD simulations and this technique conveniently allows us to take the substrate effects into consideration. We calculate the T/B ratio of dipole emission from $MoS_2$ on sapphire substrate to be 0.8 in the wavelength range of interest (FDTD Solutions, Lumerical Inc.). The top and bottom emitted power values were obtained via power monitors with sizes corresponding to the numerical apertures of the experimental setup (top NA: 0.95, bottom NA: 0.55). Then, we normalized our experimental T/B ratio map such that the bare $MoS_2$ emission T/B ratio matches this value. Therefore, the resulting T/B ratio on the NW of more than 20 takes the asymmetry of the system into consideration and is purely thanks to the antenna's directionality enhancement. Hence, it can be concluded that 25 times T/B ratio is the figure of merit directionality value of the antenna performance.

**Modified Mie theory analysis of T/B ratio.** T/B ratio values reported in Fig. 3a, are obtained as:

$$T/B = \frac{\int_{\theta=108.2°}^{\theta=251.8°} I_{ff}(\theta)d\theta}{\int_{\theta=326.5°}^{\theta=33.5°} I_{ff}(\theta)d\theta} \quad . \quad (3)$$

The integration limits were determined such that the calculated values correspond to the experiments with the correct numerical apertures: top NA = 0.95 and bottom NA = 0.55. Note that this calculation is for the emitter-antenna system in free-space without a substrate. In order to confirm that the presence of the substrate is insignificant, we obtained the same parameter sweep result of T/B ratio from FDTD simulations with a sapphire substrate (see Supplementary Fig. S4). The comparison between the FDTD result involving the substrate (Fig. S4) and the analytical result (Fig. 3a) shows that the presence of the substrate is indeed not critical for the directionality in this system.

The significantly higher T/B ratios in calculations (and simulations in Suppl. Info.) as compared to experiments are expected as the calculations only consider the emission from one source dipole placed directly underneath the NW at an optimized spacing. In experiments, however, there will be a distribution of emitters from the $MoS_2$ region underneath the NW, which results in a deviation of the emitter-to-NW distance from the optimum.

To further verify the validity of the 2D model and the simulations reported in Suppl. Info., we also conducted a few exemplary 3D simulations showing the accuracy of the 2D modeling of the considered system NW emitter system (see Figure S2).

## Acknowledgements

This research was conducted with the support from Air Force Office of Scientific Research (AFOSR) and support of the Quantum Metaphotonics and Metamaterials MURI (AFOSR Award No. FA9550-12-1-0488) and Stanford Electrical Engineering Departmental Fellowship. S.R. acknowledges support by a research grant (VKR023371) from VILLUM FONDEN. A.G.C. acknowledges the support of a Marie Curie International Outgoing Fellowship.

**Author contributions**

A.F.C. and M.L.B. conceived the idea; A.F.C., A.G.C. and M.L.B. designed the experiments; A.F.C. prepared the samples and carried out the experiments; S.R. conducted the theoretical calculations; A.F.C. conducted full-field simulations. P.K. provided guidance at the simulations and experiments. All authors analyzed and discussed the results and were involved in writing the manuscript.

# Supplementary Information

**Silicon Nanoantennas for Highly Directional Light Emission from monolayer MoS$_2$**

Ahmet Fatih Cihan[1], Alberto G. Curto[1,2], Søren Raza[1], Pieter G. Kik[1,3]

and Mark L. Brongersma[1]

4. Geballe Laboratory for Advanced Materials, Stanford University, Stanford, California 94305, United States
5. Dep. Applied Physics and Institute for Photonic Integration, Eindhoven University of Technology, 5600 MB Eindhoven, The Netherlands
6. CREOL, The College of Optics and Photonics, University of Central Florida, Florida 32816, United States

*Corresponding Author: brongersma@stanford.edu

This supplementary information consists of 6 sections in support of the main body of the text. The contents of the sections are listed below:

**S1. Comparison of Plane Wave Excitation to Far-Away Dipole Excitation**

**S2. Derivation of Scattering Efficiency of a Nanowire under Dipole Excitation**

**S3. Derivation of Far-Field Radiation Patterns**

**S4. Comparison of Radiation Patterns for 2D and 3D Simulations**

**S5. Top and Bottom Fluorescence Image Cross-Sections for NW$_1$**

**S6. FDTD Simulations of T/B Ratio**

**Section S1: Comparison of Plane Wave Excitation to Distant Dipole Excitation**

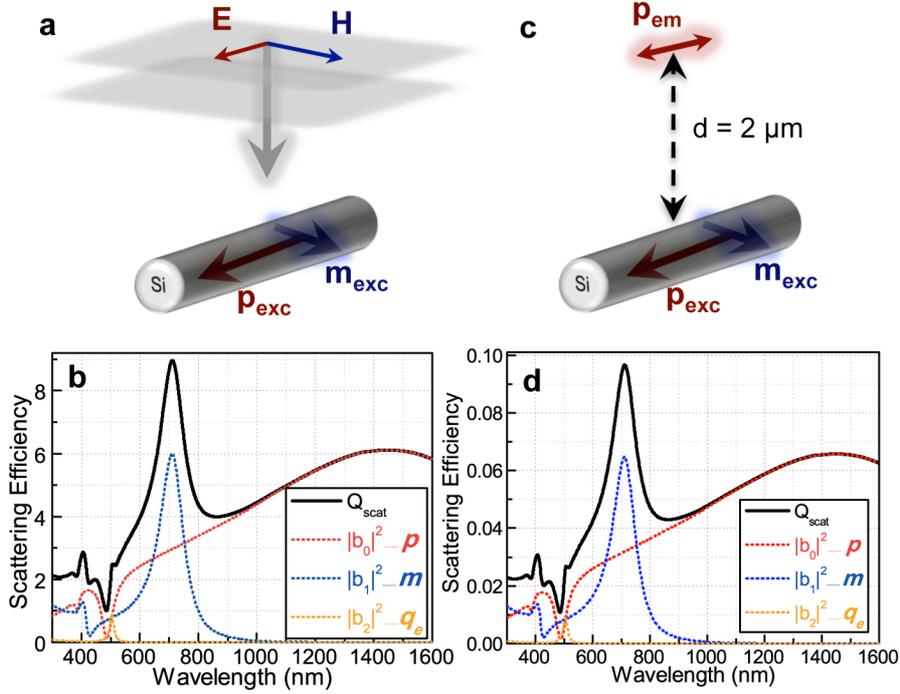

*Figure S1. Scattering from a Si nanowire (NW) for plane wave and distant dipole sources. **a**, Schematic of the NW under plane wave excitation. **b**, Scattering efficiency, $Q_{scat}$, and the modal contributions from the electric dipole, p, magnetic dipole, m, and electric quadrupole, $q_e$, under plane wave excitation for a NW with 70 nm radius. **c**, Schematic of the dipole-NW system with dipole-to-NW distance of 2 μm. **d**, Scattering efficiency and the modal contributions for the dipole case in part c.*

We compare the scattering behaviors of a Si nanowire (NW) under plane wave and dipole excitations in Fig. S1. When the dipole-to-NW distance is much larger than the excitation wavelength, we find that the spectral dependences of the scattering efficiencies become identical. However, the amplitude of the scattering efficiency is much lower in the dipole case because the NW is far away from the source, allowing only a small fraction of the emitted light to excite the multipolar resonances of the NW.

**Section S2: Derivation of Scattering Efficiency of a Nanowire under Dipole Excitation**

We present our analytical approach to calculate the electromagnetic fields of a cylindrical nanowire excited by an electric line-dipole. From the fields, we determine the dipole scattering efficiency as $Q_{\text{scat}}^{dipole} = P_{\text{scat}}/P_{\text{inc}}$ where $P_{\text{scat}}$ and $P_{\text{inc}}$ are the scattered and incident powers, respectively. We consider an infinite cylindrical NW with radius $R$ and permittivity $\varepsilon(\omega)$ placed in vacuum with the material information taken from Ref. 1. The NW is parallel to the $z$-axis and centered at the origin of the coordinate system. The dipole excitation source has a dipole moment oriented parallel to the NW and is positioned at $x = r_s = R + d$, where $d$ is the dipole-to-NW distance. For this dipole orientation, the source is equivalent to a line current source.

The incident electric field of the dipole can be expressed as [in cylindrical coordinates (r, θ, z)]:[2]

$$\mathbf{E^i} = \hat{\mathbf{z}} E_z^i = \hat{\mathbf{z}} K H_0^{(1)}(k_0|\mathbf{r} - \mathbf{r_s}|) = \hat{\mathbf{z}} K \begin{cases} \sum_{n=-\infty}^{\infty} J_n(k_0 r) H_n^{(1)}(k_0 r_s) e^{-in\theta}, \text{ for } r \leq r_s \\ \sum_{n=-\infty}^{\infty} J_n(k_0 r_s) H_n^{(1)}(k_0 r) e^{-in\theta}, \text{ for } r \geq r_s \end{cases}$$

where $k_0 = 2\pi/\lambda$ is the vacuum wave number, $H_n^{(1)}$ is the Hankel function of first order, $J_n$ is the Bessel function, and $K = \omega\mu_0 I/4$ where I is the line current.[2]

The electric fields scattered by and inside the NW can be expanded as

$$\mathbf{E^s} = \hat{\mathbf{z}} E_z^s = -\hat{\mathbf{z}} K \sum_{n=-\infty}^{\infty} b_n H_n^{(1)}(k_0 r) H_n^{(1)}(k_0 r_s) e^{-in\theta}, \text{ for } r \geq R.$$

$$\mathbf{E^c} = \hat{\mathbf{z}} E_z^c = \hat{\mathbf{z}} K \sum_{n=-\infty}^{\infty} c_n J_n(kr) H_n^{(1)}(k_0 r_s) e^{-in\theta}, \quad \text{for } r \leq R.$$

respectively, where $k = k_0\sqrt{\varepsilon(\omega)}$ is the wave number in the NW. Applying Maxwell's boundary conditions at the surface of the NW determines the expansion coefficients $b_n$ and $c_n$. We find $b_n$ to be the conventional Mie coefficient also found in the case of plane-wave excitation [3]

$$b_n = \frac{J_n(kR)J_n'(k_0R) - \sqrt{\varepsilon(\omega)}J_n(k_0R)J_n'(kR)}{J_n(kR)H_n^{(1)'}(k_0R) - \sqrt{\varepsilon(\omega)}H_n^{(1)}(k_0R)J_n'(kR)}.$$

Now that we have the scattered and incident electric fields, we can calculate the scattered and incident powers:[4]

$$P_{\text{scat}} = R\int_0^{2\pi} d\theta\, \mathbf{S}_{\text{scat}}(r = R, \theta) \cdot \hat{\mathbf{r}},$$

where $\mathbf{S}_{\text{scat}} \cdot \hat{\mathbf{r}}$ is the radial component of the Poynting vector associated with scattered fields. $\mathbf{S}_{\text{scat}} = \frac{1}{2}Re\{\mathbf{E}^s \times \mathbf{H}^{s*}\}$. After some straightforward but lengthy algebra we arrive at the result

$$P_{\text{scat}} = \frac{1}{8}\omega\mu_0 I^2 \sum_{n=-\infty}^{\infty} |b_n|^2 \left|H_n^{(1)}[k_0(R+d)]\right|^2.$$

Similarly, the incident power can be derived as

$$P_{\text{inc}} = \frac{1}{8}\omega\mu_0 I^2.$$

Therefore, the scattering efficiency is

$$Q_{\text{scat}}^{\text{dipole}} = \frac{P_{\text{scat}}}{P_{\text{inc}}} = \sum_{n=-\infty}^{\infty} |b_n|^2 \left|H_n^{(1)}[k_0(R+d)]\right|^2.$$

For comparison, we add that the scattering efficiency for the plane wave case is

$$Q_{\text{scat}}^{\text{plane-wave}} = \frac{2}{k_0 R} \sum_{n=-\infty}^{\infty} |b_n|^2$$

which is independent of the source position unlike the dipole case.

**Section S3: Derivation of Far-Field Radiation Patterns**

To derive the far-field radiation pattern, we must examine the equations for the incident and scattered field for large distance $r$. Utilizing the asymptotic relations for the Hankel functions of first order

$$\lim_{k_0 r \to \infty} H_n^{(1)}(k_0 r) = \sqrt{\frac{2}{\pi}} (-i)^n e^{-i\pi/4} \frac{e^{ik_0 r}}{\sqrt{k_0 r}},$$

we determine the incident and scattered far-fields as

$$\lim_{k_0 r \to \infty} \mathbf{E}^i = \hat{\mathbf{z}} K \sqrt{\frac{2}{\pi}} e^{-i\pi/4} \sum_{n=-\infty}^{\infty} (-i)^n J_n(k_0 r_s) e^{-in\theta} \frac{e^{ik_0 r}}{\sqrt{k_0 r}} \equiv \mathbf{E}_{\text{ff}}^i(\theta) \frac{e^{ik_0 r}}{\sqrt{k_0 r}}$$

$$\lim_{k_0 r \to \infty} \mathbf{E}^s = -\hat{\mathbf{z}} K \sqrt{\frac{2}{\pi}} e^{-i\pi/4} \sum_{n=-\infty}^{\infty} (-i)^n b_n H_n^{(1)}(k_0 r_s) e^{-in\theta} \frac{e^{ik_0 r}}{\sqrt{k_0 r}} \equiv \mathbf{E}_{\text{ff}}^s(\theta) \frac{e^{ik_0 r}}{\sqrt{k_0 r}}.$$

Hence, the total electric far-field is

$$\mathbf{E}_{\text{ff}}^t(\theta) = \mathbf{E}_{\text{ff}}^i(\theta) + \mathbf{E}_{\text{ff}}^s(\theta) = \hat{\mathbf{z}} K \sqrt{\frac{2}{\pi}} e^{-i\pi/4} \sum_{n=-\infty}^{\infty} (-i)^n \left[ J_n(k_0 r_s) - b_n H_n^{(1)}(k_0 r_s) \right],$$

and the far-field intensity becomes:

$$I_{\text{ff}}(\theta) = |\mathbf{E}_{\text{ff}}^t(\theta)|^2 = \frac{1}{8\pi} \omega^2 \mu_0^2 I^2 \left| \sum_{n=-\infty}^{\infty} (-i)^n e^{-in\theta} \left[ J_n(k_0 r_s) - b_n H_n^{(1)}(k_0 r_s) \right] \right|^2.$$

**Section S4: Comparison of Radiation Patterns for 2D and 3D Simulations**

To show the validity of the line-dipole-based analytical approach, we conduct 2D and 3D Finite-Difference-Time Domain (FDTD) simulations which correspond to line and point dipole sources, respectively. Figure S2 shows far-field radiation patterns for two extreme examples: high T/B and high B/T ratios. In the plane orthogonal to the NW and intersecting the emitter, which is relevant for the experiments, the 2D results closely resemble the 3D results. as would also be intuitively expected to be originating from symmetry argument. Considering that our experimental spectral region of interest is around 680 nm, we can say that 2D model is working well for our analysis.

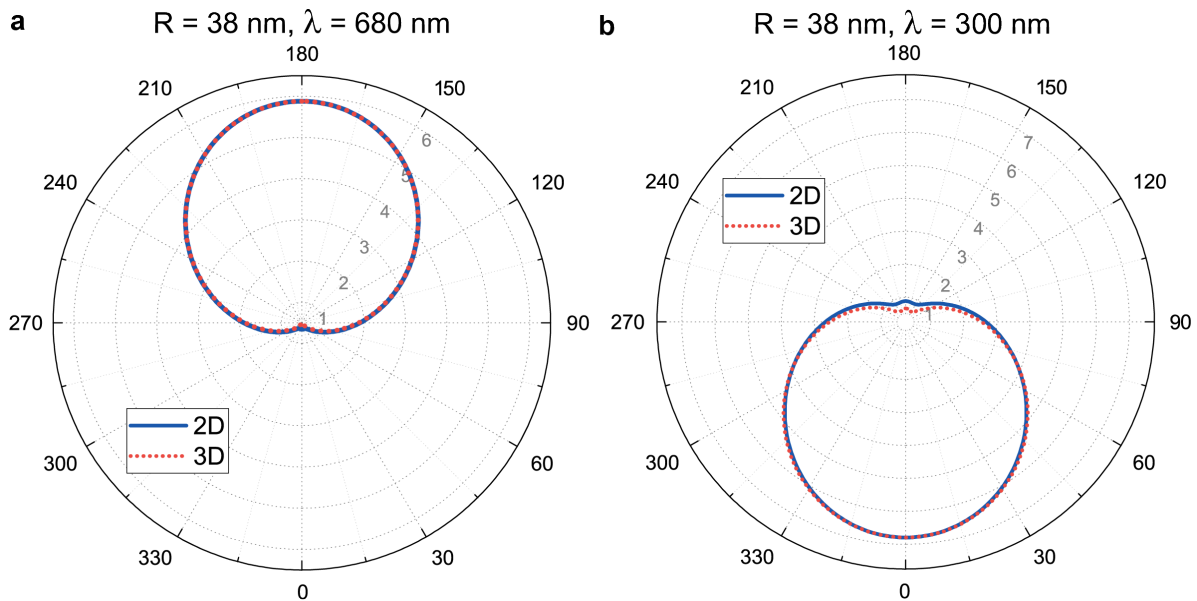

*Figure S2. Comparison of radiation patterns for 2D and 3D FDTD simulations. Angular radiation patterns on the plane perpendicular to the NW and intersecting the emitter from 2D and 3D simulations for the wavelength of **a**, 680 nm and **b**, 300 nm. The NW radius is 38 nm and the polarization is TM.*

**Section S5: Top and Bottom Fluorescence Image Cross-Sections for NW$_1$**

Figure S3 shows images and linecuts of the fluorescence emission from NW$_1$ described in the main text, whose radius is tapered from 20 – 40 nm. The data shows that the directionality of the MoS$_2$ emission is much easier to observe after etching the MoS$_2$ layer with the Si nanowire used as a mask. Note that in Figure S3h, the bottom emission across the detection line is at the noise level while the top emission is remaining. This indicates that the NW almost completely suppresses the bottom emission and directs virtually all of the emission in the top direction.

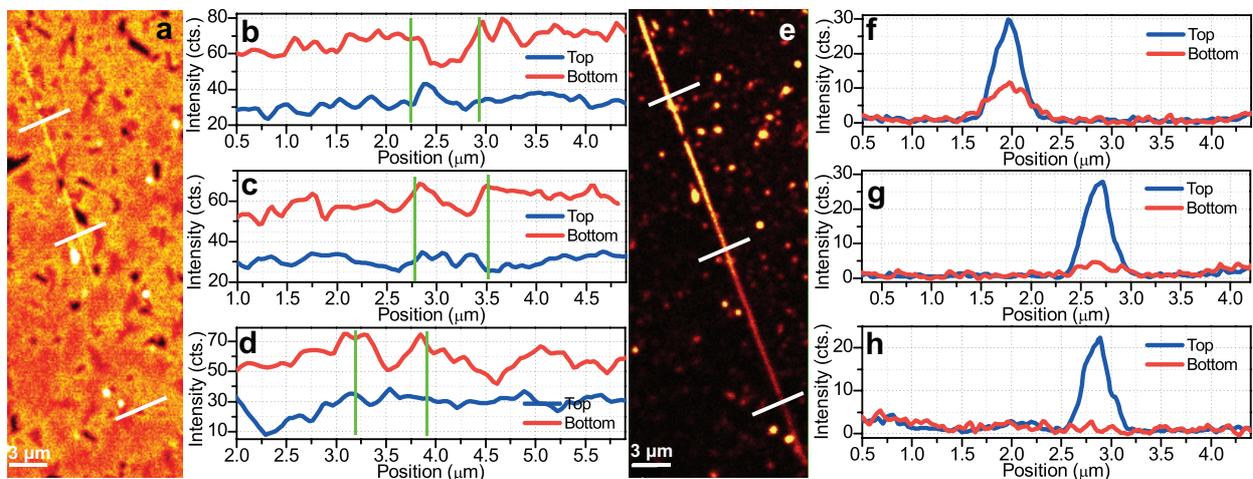

*Figure S3. Line cross-sections of fluorescence images of the sample before and after etching. a, Top fluorescence image of the sample before etching. b-d, Cross-sections of the top and bottom images of the sample before etching taken along the white lines shown in part a. The green lines indicate the position of the NW to guide the eye. e, Top fluorescence image of the sample after etching. f-h, Cross-sections after the etching procedure.*

**Section S6: FDTD Simulations of T/B Ratio**

In order to further verify the validity of our modified Mie theory approach, we conducted a series of Finite Difference Time Domain (FDTD) simulations. In these 2D full-field simulations we also included the sapphire substrate to check the validity of the analytical result reported in Fig. 3a of the main text. Since the T/B ratio result shown in Fig. S4 is consistent with Fig. 3a, we can say that the substrate has negligible impact on the directionality behavior of our emitter-antenna system.

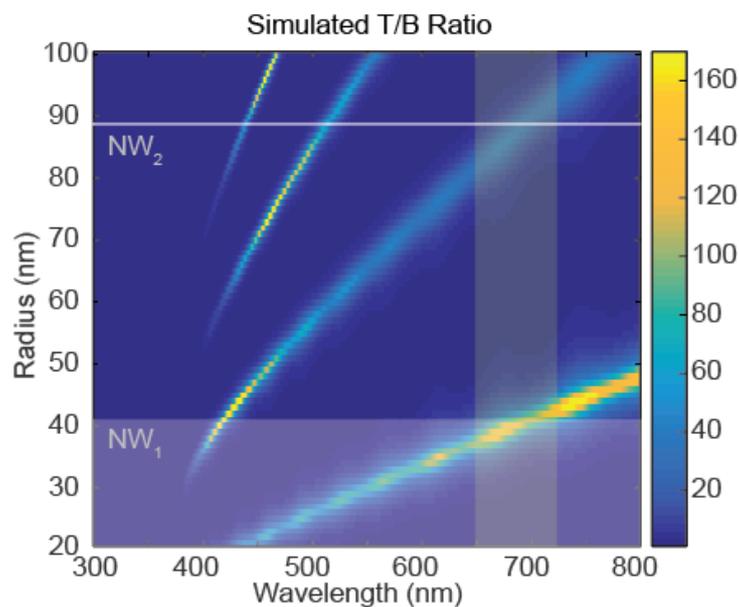

*Figure S4. Numerical confirmation of the validity of modified Mie theory approach. T/B ratio obtained with 2D FDTD simulations. The simulated structure includes a sapphire substrate and the T/B ratio is obtained via power monitors whose sizes are determined based on the experimental numerical apertures (0.95 for top and 0.55 for bottom).*

## Section S7: Supplementary References